\documentclass{emulateapj}
\usepackage{apjfonts}
\usepackage{natbib}

\newcommand{\src}{GS~1826-24}

\shorttitle{X-ray Bursts from \src}
\shortauthors{Heger et al.}

\begin{document}

\title{Models of Type I X-ray Bursts from \src: A Probe of rp-Process Hydrogen Burning}

\author{Alexander Heger\altaffilmark{1}, Andrew Cumming\altaffilmark{2},
Duncan K.~Galloway\altaffilmark{3,4}, and Stanford E.~Woosley\altaffilmark{5}}

\altaffiltext{1}{Theoretical Astrophysics Group, T-6, MS B227, Los Alamos National
Laboratory, Los Alamos, NM 87545; 1@2sn.org}
\altaffiltext{2}{Department of Physics, McGill University, Montreal QC, H3A 2T8, Canada}
\altaffiltext{3}{School of Physics, University of Melbourne, Victoria, Australia}
\altaffiltext{4}{present address: School of Physics \& School of Mathematical Sciences, Monash University, Clayton 3800, Australia}
\altaffiltext{5}{Department of Astronomy and Astrophysics, University of California, Santa Cruz, CA 95064}

\begin{abstract}
The X-ray burster \src\  shows extremely regular Type I X-ray bursts whose energetics and recurrence times agree well with thermonuclear ignition models. We present calculations of sequences of burst lightcurves using  multizone models which follow the nucleosynthesis ($\alpha$p and rp-processes) with an extensive nuclear reaction network. The theoretical and observed burst lightcurves show remarkable agreement. The models naturally explain the slow rise (duration $\approx 5\ {\rm s}$) and long tails ($\approx 100\ {\rm s}$) of these bursts, as well as their dependence on mass accretion rate. This comparison provides further evidence for solar metallicity in the accreted material in this source, and gives a distance to the source of $(6.07\pm 0.18)\ {\rm kpc}\ \xi_b^{-1/2}$, where $\xi_b$ is the burst emission anisotropy factor. The main difference is that the observed lightcurves do not show the distinct two-stage rise of the models. This may reflect the time for burning to spread over the stellar surface, or may indicate that our treatment of heat transport or nuclear physics needs to be revised. The trends in burst properties with accretion rate are well-reproduced by our spherically symmetric models which include chemical and thermal inertia from the ashes of previous bursts. Changes in the covering fraction of the accreted fuel are not required.
\end{abstract}


\section{Introduction}

The basic physics of Type I X-ray bursts as thin shell flashes on the surfaces of accreting neutron stars was understood many years ago (e.g.~\citealt{fuj81,LE95}). Nonetheless, detailed comparisons of observations and theory are often less than successful \citep{fuj87,vplewin}. The burster \src\ (aka Ginga 1826-24) is an important exception. \cite{Ubertini99} dubbed it the ``clocked burster'' because of its extremely regular bursting behavior. They found a burst recurrence time close to 6 hours with a dispersion of approximately 6 minutes. \cite{Bildsten00} dubbed it the ``textbook burster'' because of the good agreement with theory. He noted that the burst energetics and recurrence times were as expected for the inferred accretion rate $\dot M\approx 10^{-9}\ M_\odot\ {\rm yr^{-1}}$, and proposed that the long burst tails, lasting $\approx 100\ {\rm s}$, were powered by rp-process hydrogen burning \citep{WW81,HF84}. The rp-process involves a series of proton captures and beta-decays on heavy nuclei close to the proton drip line, for which there are considerable uncertainties in beta decay rates, reaction rates, and nuclear masses. Type I X-ray bursts offer an important test of our understanding of this process.

\cite{gal04} (hereafter G04) studied 24 bursts observed from \src\ by the {\em Rossi X-ray Timing Explorer (RXTE)} between 1997 November and 2002 July, and carried out a detailed comparison of the observed recurrence times and energetics with theory. During this period, the accretion rate (assumed to be proportional to the persistent X-ray luminosity) increased by $\approx 50$\%, while the burst recurrence time decreased from $\Delta t\approx 6$ hours to $\approx 4$ hours. At the observed accretion rate $\dot M\approx 10^{-9}\ M_\odot\ {\rm yr^{-1}}$, hydrogen burns stably between bursts by the beta-limited hot CNO cycle \citep{HF65}, heating the accumulating layer of hydrogen and helium at a rate that depends only on the mass fraction of CNO elements. The resulting  ignition column depth of $\approx 2\times 10^8\ {\rm g\ cm^{-2}}$ is nearly independent of $\dot M$ (e.g.~\citealt{CB00}), in agreement with the observed scaling which is close to $\Delta t\propto 1/\dot M$.

G04 used the variation of burst energetics with accretion rate to constrain the composition of the accreted material. The energetics are measured by the ratio $\alpha$ of integrated persistent flux between bursts to the burst fluence, equivalent to the ratio of the gravitational energy release from accretion to the nuclear energy release in bursts. The observed value $\alpha\approx 40$ is close to the value expected for the $4$--$5\ {\rm MeV}$ per nucleon energy release in hydrogen burning \citep{Bildsten00}. G04 found that $\alpha$ decreased from $\approx 44$ in 1997 to $\approx 40$ in 2002. This  is consistent with the change in composition of the fuel layer expected from hot CNO burning between bursts, if the accreted material has solar metallicity. For a shorter recurrence time, less hydrogen burns between bursts, resulting in a larger burst energy and smaller $\alpha$. In addition, G04 found that whereas the ignition models predicted that the ignition column depth should increase slightly with $\dot M$, the observed trend was the opposite. 

The study of G04 did not include models of the burst lightcurves. In this paper, we compare observations of \src\ with multizone models \citep{woo04}, which include a large nuclear reaction network to follow the rp-process in detail. We show that these models naturally explain the slow rise (duration $\approx 5\ {\rm s}$) and long tails ($\approx 100\ {\rm s}$) of these bursts. We show that our time-dependent models reproduce the scalings of the recurrence time with accretion rate for a metallicity near solar.

\begin{figure}
\epsscale{1.6}
\plotone{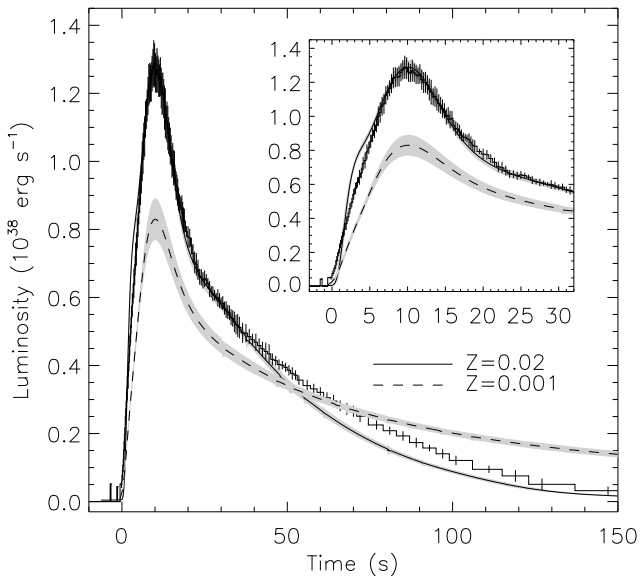}
\caption{Comparison of observed and calculated lightcurves. The histogram shows the average lightcurve from the bursts observed during the year 2000 when the recurrence time was $\approx 4$ hours (G04, Figure 2). The error bars are the 1$\sigma$ variations from burst to burst. The solid and dashed curves are the average burst profiles from models A3 ($Z=0.02$) and B3 ($Z=0.001$), which have $\Delta t=3.9$ and $4.0$ hours respectively. 
The inset magnifies the rise and the initial part of the decay.
The grey bands indicate the 1$\sigma$ variation of the burst profiles about the average.}
\label{fig:lc}
\end{figure}

\section{Comparison of Observations with Time-dependent Burst Models}

We calculate sequences of bursts as described by \cite{woo04} (hereafter W04). The KEPLER code \citep{Wea78} is used to follow the evolution of the outer layers of the neutron star through a sequence of successive bursts. A large adaptive nuclear reaction network is used to follow the nucleosynthesis at each depth, and we include convection when needed using time-dependent mixing length theory. We use the same input nuclear physics, stellar opacities, and neutron star parameters as W04, but consider a wider range of accretion rates. The Newtonian calculations are corrected for general relativity as described in \S 4.4 of W04 for a neutron star mass of $1.4\ M_\odot$. The corresponding stellar radius is $R=11.2\ {\rm km}$, and gravitational redshift is $z=0.26$.

A summary of the results is given in Table \ref{tab:summary}. For each sequence of bursts, we list the rest mass accretion rate $\dot M$, recurrence time $\Delta t$, burst energy $E_{\rm burst}$, gravitational mass accumulated between bursts $\Delta M=\dot M\Delta t/(1+z)$, and $\alpha=F_X\Delta t/E_{\rm burst}=\Delta Mc^2z/E_{\rm burst}$, all as seen by an observer at infinity. The quantities given are averaged over all bursts except the first burst in each sequence, which is typically more energetic than the subsequent bursts (W04). We give the standard deviation of each quantity in parentheses, to show the burst to burst variations.
Models A4 and B3, which have $\dot M=1.75\times 10^{-9}\ M_\odot\ {\rm yr^{-1}}$, have the same parameters as models ZM and zM of W04. There are slight differences at the level of $\approx 3$\% between the burst properties in models A4 and B3 as compared to models ZM and zM of W04, because of refinements of the KEPLER code that were made following publication of the W04 paper.

We compare these simulations to bursts observed by {\it RXTE}\/ between 1997 November and 2002 July. We analyze the data as described in G04, with the following exceptions: 1) the spectral fitting was performed using {\sc lheasoft} version 5.3, released 2003 November 17, for which the effective area of the proportional counter instrument (and hence the source flux) was reduced by approximately 15\% compared to earlier versions; 2) improved calculation of the burst fluence which was better able to handle gaps in the data, which increased the estimated fluence in some cases by at most 5\%. These changes also had the effect of reducing the absolute $\alpha$-values, although the trend with persistent flux was unchanged.

In Figure \ref{fig:lc}, we compare the mean lightcurve for bursts observed during 2000 (G04 Figure 2) with the mean burst lightcurves from models A3 and B3. These models are chosen because they have similar recurrence times to the observed recurrence time of $4.1$ hours. We calculate the mean lightcurves by aligning bursts in each sequence by their peak luminosities.  The error bars in Figure \ref{fig:lc} show the 1$\sigma$ burst to burst variation about the mean observed lightcurve. The shaded region shows the same variation for the theoretical lightcurves.

For this comparison, we choose the distance to the source (within the allowed range $4<d<8$ kpc; G04) so that the peak luminosity of the observed bursts agrees with the peak luminosity of bursts from model A3. The relation between the peak burst luminosity $L_{\rm peak}$ and the observed peak flux is $4\pi d^2\xi_bF_{\rm peak}=L_{\rm peak}$, where $\xi_b$ is a factor that accounts for possible anisotropy in the burst emission (e.g.~Fujimoto 1988). The average observed peak flux in the 2000 epoch is $(2.93\pm 0.15)\times 10^{-8}\ {\rm erg\ cm^{-2}\ s^{-1}}$, and the average peak luminosity of bursts in model A3 is $L_{\rm peak}=(1.29\pm 0.04)\times 10^{38}\ {\rm erg\ s^{-1}}$, giving a distance $d=(6.07\pm 0.18)\ {\rm kpc}\ \xi_b^{-1/2}$.

\begin{deluxetable*}{lccccccc}
\tablecaption{Average Burst Properties\label{tab:summary}\tablenotemark{a}}
\tablewidth{0pt}
\tablehead{
\colhead{Model} & \colhead{Number} & \colhead{Z} & \colhead{$\dot M$} & \colhead{$\Delta t$} & 
\colhead{$E_{\rm burst}$} & \colhead{$\alpha$} & \colhead{$\Delta M$} \nl
\colhead{} & \colhead{of bursts} & \colhead{} & \colhead{($10^{-9}\ M_\odot\ {\rm yr^{-1}}$)} & \colhead{(h)} &
\colhead{($10^{39}\ {\rm ergs}$)} & \colhead{} & \colhead{($10^{21}\ {\rm g}$)}\nl
}
\startdata 
A1 & 19 & 0.02 & 1.17 & 5.4 (0.1) & 4.67 (0.20) & 57.4 (2.8) & 1.14 (0.03)\nl
A2 & 18 & 0.02 & 1.43 & 4.3 (0.1) & 4.67 (0.11) & 55.6 (1.2) & 1.11 (0.03)\nl
A3 & 30 & 0.02 & 1.58 & 3.85 (0.06) & 4.73 (0.07) & 55.0 (0.9) & 1.10 (0.02)\nl
A4 & 13 & 0.02 & 1.75 & 3.48 (0.06) & 4.84 (0.06) & 53.6 (0.8) & 1.11 (0.02)\nl
B1 & 12 & 0.001 & 1.17 & 12.8 (0.6) & 13.3 (0.7) & 47.8 (0.4) & 2.71 (0.14)\nl
B2  & 17 & 0.001 & 1.43 & 6.04 (0.41) & 7.74 (0.49) & 47.4 (1.0) & 1.57 (0.11)\nl
B3  & 15 & 0.001 & 1.75 & 3.98 (0.28) & 6.26 (0.32) & 47.3 (2.4) & 1.27 (0.09)\nl
\enddata
\tablenotetext{a}{See text for definitions of quantities.}
\end{deluxetable*}

\begin{figure}
\epsscale{1.2}
\plotone{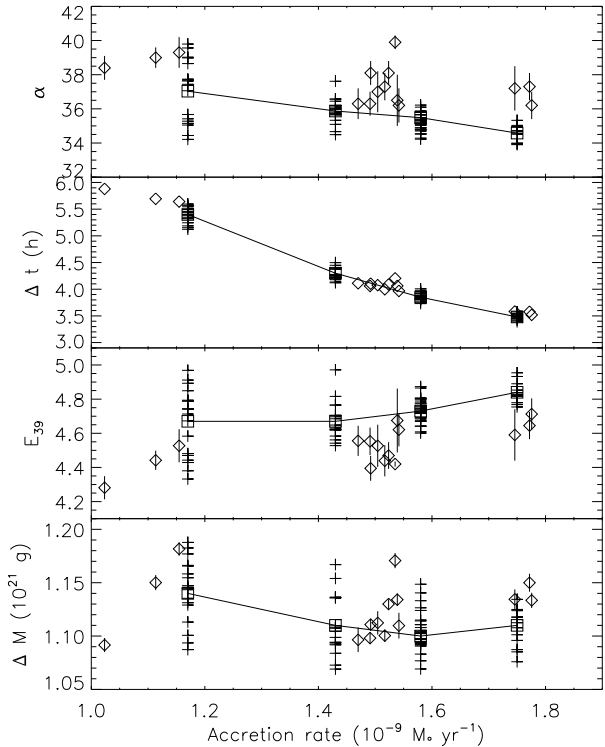}
\caption{Burst parameters for the $Z=0.02$ models A1--A4
compared with individual observed bursts. The properties of the observed bursts are plotted as diamonds. The properties of individual model bursts are plotted as crosses, with average values for each model plotted as squares joined by solid lines. 
}
\label{fig:alpha}
\end{figure}

Once the bursts have been normalized in this way, the agreement between the observed and theoretical lightcurves for $Z=0.02$ (model A3) is remarkable. Model A3 fits the observed decay exceedingly well out to about 40s, falling a little below the observed flux between 40 and 120s. The burst to burst variations in the models are also of comparable magnitude to the burst to burst variations in the data. 
The most significant difference is that the theoretical model shows a distinct two-stage rise which is not apparent in the observed lightcurve (see the lower panel of Fig.~1).

Model B3, which has a low metallicity, does not reproduce the observed lightcurve. Given the uncertainty in the distance to the source, the normalization may be adjusted to bring the observed and predicted peak luminosities into agreement, but the shape of the decay provides an additional constraint. In model B3, the lower metallicity leads to very little hydrogen burning between bursts, giving less helium and more hydrogen at ignition than in the solar metallicity model. The result is a  burst with a much longer tail, inconsistent with the observed profile for any distance.

\begin{figure}
\epsscale{1.6}
\plotone{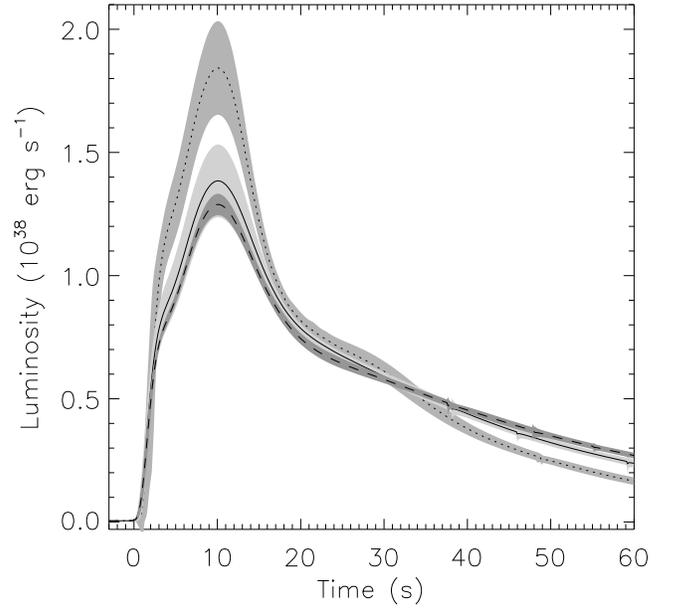}
\caption{Change in the lightcurve as a function of accretion rate. Averaged lightcurves from models A1 (dotted curve), A2 (solid curve), and A3 (dashed curve). The grey shaded areas indicate the standard deviation of the burst to burst variation  around the average lightcurve. As accretion rate increases, the burst becomes less luminous, with a longer rise time and slower decay.}
\label{fig:lc3}
\end{figure}

\begin{figure}
\epsscale{1.6}
\plotone{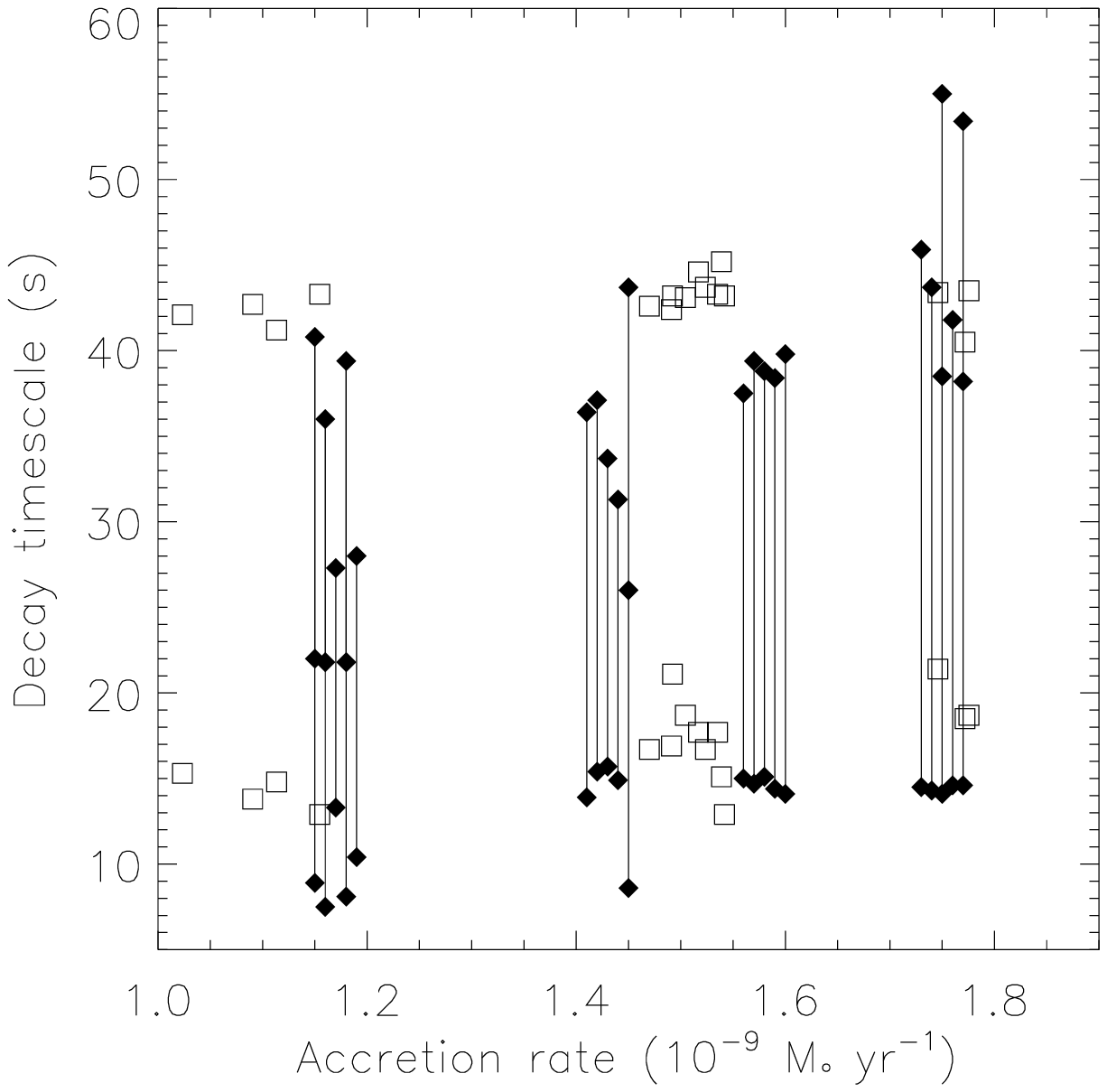}
\caption{Fitted exponential decay timescales for observed bursts (open symbols) and theoretical lightcurves from models A1, A2, A3 and A4 (solid symbols). For the models, we choose a representative sample of 5 bursts (burst numbers 7 to 11 from each run), and displace them slightly on the horizontal axis for clarity. Note that some of the theoretical lightcurves show a three stage decay. The uncertainties in the fitted decay times are comparable to the scatter in the points. We adopt the same relation between $F_X$ and $\dot M$ as in Figure \ref{fig:alpha}.}
\label{fig:times}
\end{figure}

In Figure \ref{fig:alpha}, we compare the burst recurrence times, energies, $\alpha$ values, and ignition masses as a function of accretion rate for the observed bursts and the solar metallicity models A1--A4. The observed burst fluence $E_b$ is related to the burst energy $E_{\rm burst}$ by $E_b=4\pi d^2\xi_bE_{\rm burst}$, and we adopt the same value of $\xi_bd^2$ that we determined earlier by comparing the burst peak luminosity with the observed peak flux. We convert the bolometric flux $F_X$ to accretion rate $\dot M$ using the relation $4\pi d^2\xi_pF_X=\dot Mc^2 z/(1+z)$, where $\xi_p$ is a factor that accounts for possible anisotropy in the persistent emission (e.g.~Fujimoto 1988). For the same value of distance, we find that $\xi_p/\xi_b=1.55$ is required for the recurrence times to match in Figure \ref{fig:alpha}. The theoretical alpha values have been divided by this same factor to give the correct prediction for alpha including anisotropic emission.

The agreement between the models and observations in Figure \ref{fig:alpha} is excellent. The main difference is that the burst energy is overpredicted by approximately 5\%, and the $\alpha$ value underpredicted by the same factor. The low metallicity models B1--B3 are not consistent with the observed trends. For example, the burst energies show a factor of 2 variation in these models over the observed accretion rate range (Table 1), much larger than the $\approx 10$\% variations seen in the data.

Figure \ref{fig:lc3} shows how the burst lightcurves change with accretion rate (models A1 to A3). G04 pointed out that the burst lightcurves changed with recurrence time. They fitted the decay of the bursts to an exponential profile with a break in the exponential decay time. The theoretical curves show a similar two timescale decay (this is particularly apparent in the lower panel of Fig.~1). G04 noted a significant increase in the first exponential decay timescale as $\Delta t$ decreased, from $14.7\pm 0.7$s when $\Delta t\approx 6$ hours to $17.5\pm 1.1$s when $\Delta t\approx 4$ hours. The second exponential decay timescale $43\pm 1$s showed no significant change. G04 speculated that this change was a result of the composition of the fuel, since more hydrogen is depleted between bursts for larger $\Delta t$.

Figure \ref{fig:lc3} shows that the model lightcurves show a similar change with accretion rate to the observed lightcurves, presumably relating to the changing hydrogen/helium ratio at ignition. This is quantified in Figure \ref{fig:times}, which shows fitted exponential decay times for the observed and model bursts.  Overall, the agreement is good, although the decay times of the model bursts are generally slightly shorter than observed (consistent with the difference in the burst tails in Figure 1), and they show much larger burst to burst variations than observed. Also, whereas the observed bursts are well fit by a two-timescale decay, some of the model bursts show a more complex decay profile, which we have fitted with three timescales.

\section{Discussion}

We have presented a first comparison between multizone burst models and the observed bursts from \src. The good agreement further confirms \src\ as a ``textbook'' burster. For a distance of $(6.07\pm 0.18)\ {\rm kpc}\ \xi_b^{-1/2}$ (where $\xi_b$ is the burst emission anisotropy factor), our solar metallicity models closely match the observed lightcurves. Both the long rise and decay times arise naturally from rp-process hydrogen burning. From the very regular nature of the bursting, G04 argued that the accreted material covers the entire surface of the neutron star; this is supported by the excellent agreement of our spherically-symmetric models. Solar metallicity models agree best with the observed lightcurves. Low metallicity models produce too little  helium by hot CNO burning prior to ignition, leading to a lower peak luminosity and a longer rp-process tail. This agrees with the conclusions of G04, who argued for solar metallicity based on the burst energetics. The estimate of the distance is based on a comparison of mean burst lightcurve at a single epoch, with the lightcurve predicted by the model for parameters giving the same recurrence time. It is possible that lightcurve comparisons at different epochs (i.e. values of the recurrence time) may result in different values of the distance and/or anisotropy parameters. Such comparisons will provide an additional consistency check for the distance estimation, which we will undertake in a future paper.

Our models reproduce the observed variation in burst recurrence times, energies, and $\alpha$ values  with accretion rate for a ratio of anisotropy factors for the persistent and burst emission of $\xi_p/\xi_b=1.55$. This is within the range discussed by \cite{fuji88} and similar to that inferred for other sources (e.g.~\citealt{sztajno}). G04 found that the ignition models of \cite{CB00} reproduced the change in $\alpha$, but not the scaling of ignition column with $\dot M$.  In these models, there is less time for helium production between bursts at higher $\dot M$, delaying the ignition, and leading to increasing ignition mass. However, our time-dependent models show that the chemical and thermal inertia associated with the ashes from previous bursts (Taam 1980) outweighs the lower helium abundance, heating the layer, and leading to a {\em smaller} ignition mass as $\dot M$ increases, in agreement with observations. A change in the covering fraction of accreted fuel, as speculated by G04, is not required.

Despite the good agreement, there are several differences between the observations and the models which need to be investigated in future work. These comparisons promise to constrain the nuclear physics of the rp-process, the thermal properties of the burning layers, spreading of the nuclear burning, and neutron star parameters. The model burst lightcurves show a distinct two-component rise which is not present in the data, as well as a slightly shorter burst tail. Another difference in burst shape is that, unlike the observations, some of the model lightcurves show a three-stage rather than two-stage decay. These differences may relate to our treatment of heat transport (especially convection) or nuclear physics. 
W04 confirmed previous findings that burst tails are sensitive to nuclear flows above the iron group \citep{sch01a,koi99}, and also pointed out that the rise times are sensitive to nuclear decays \emph{below} the iron group. An alternative for the rise is that a finite propagation time for the burning around the star of $\sim 1\ {\rm s}$ (e.g.~\citealt{fryxellwoosley}; \citealt{spitkovsky}) might act to ``wash out'' the kink. 

The neutron star mass and radius also change the predicted burst properties. In this paper, we have considered a neutron star with redshift factor $z=0.26$, corresponding to $M=1.4\ M_\odot$ and $R=11.2\ {\rm km}$. A smaller radius of $10.6\ {\rm km}$ would reduce the predicted burst energies by 5\%, and would increase the redshift factor and therefore $\alpha$, bringing both of these quantities into better agreement with observations (Fig.~2). We will investigate the constraints on the neutron star mass and radius in detail in future work.

\acknowledgements
AH is supported by the Department of Energy under grant W-7405-ENG-36 to the Los Alamos National Laboratory. AC is grateful for support from NSERC, CIfAR, FQRNT, and is an Alfred P.~Sloan Research Fellow. SW acknowledges support from the NSF (AST 02-06111), NASA (NAG5-12036), and the DOE Program for Scientific Discovery through Advanced Computing (SciDAC; DE-FC02-01ER41176).

\end{document}